\documentclass[a4paper]{article}
\usepackage{amscd,amsmath,amssymb}

\newcommand{\R}{\mathbb R}
\newcommand{\Z}{\mathbb Z}

\newcommand{\p}[1]{(\ref{#1})}

\newcommand{\cA}{{\cal A}}

\newcommand{\bG}{{\overline G}{}}

\newcommand{\bp}{{\bar p}}

\newcommand{\bu}{{\bar u}}

\newcommand{\bv}{{\bar v}}

\newcommand{\sfrac}[2]{{\textstyle\frac{#1}{#2}}}
\newcommand{\und}{\qquad\textrm{and}\qquad}

\newcommand{\be}{\begin{equation}}
\newcommand{\ee}{\end{equation}}
\newcommand{\bea}{\begin{eqnarray}}
\newcommand{\eea}{\end{eqnarray}}

\newcommand{\ba}{\begin{array}} \newcommand{\ea}{\end{array}}

\def\im{{\rm i}}

\newcommand{\nn}{\nonumber}

\begin{document}

\pagenumbering{gobble}

\begin{flushright}
ITP-UH-16/16
\end{flushright}

\vspace{2cm}

\begin{center}
{\LARGE\bf Minimal realization of $\ell$-conformal Galilei algebra,\\[0.5cm]
Pais--Uhlenbeck oscillators and their deformation}
\end{center}

\vspace{1cm}

\begin{center}
{\Large\bf  Sergey Krivonos${}^{a}$ , Olaf Lechtenfeld${}^{b}$ and Alexander Sorin${}^{a,c,d}$
}
\end{center}

\vspace{1cm}

\begin{center}
${}^a$ {\it
Bogoliubov  Laboratory of Theoretical Physics, JINR,
141980 Dubna, Russia}

\vspace{0.2cm}

${}^b$ {\it
Institut f\"ur Theoretische Physik and Riemann Center for Geometry and Physics \\
Leibniz Universit\"at Hannover,
Appelstrasse 2, D-30167 Hannover, Germany}

\vspace{0.2cm}

${}^c$ {\it
National Research Nuclear University MEPhI (Moscow Engineering Physics Institute), \\
Kashirskoe Shosse 31, 115409 Moscow, Russia}

\vspace{0.2cm}

${}^d$ {\it
Dubna International University, 141980, Dubna,
Russia}

\vspace{0.5cm}

{\tt krivonos@theor.jinr.ru, lechtenf@itp.uni-hannover.de, sorin@theor.jinr.ru}

\end{center}

\vspace{2cm}

\begin{abstract}\noindent
We present the minimal realization of the $\ell$-conformal Galilei group in $2{+}1$ dimensions on a single complex field.
The simplest Lagrangians yield the complex Pais--Uhlenbeck oscillator equations. We introduce a minimal
deformation of the $\ell{=}1/2$ conformal Galilei (a.k.a.\ Schr\"odinger) algebra and construct the corresponding
invariant actions. Based on a new realization of the $d{=}1$ conformal group,
we find a massive extension of the near-horizon Kerr--dS/AdS metric.
\end{abstract}

\newpage
\pagenumbering{arabic}
\setcounter{page}{1}

\section{Introduction}
Besides their application to the nonrelativistic AdS/CFT correspondence, nonrelativistic conformal
algebras \cite{confNH1,henkel,confNH2} attract attention due to their non-trivial structure.
The conformal Galilei algebra is determined by a positive half-integer or integer number $\ell$
and the number~$d$ of spatial dimensions. Most of its features do not depend on the dimension, except in
the special case of~$d{=}2$, where an additional central-charge extension is admitted~\cite{exotic,gen1}.
The (more relevant) parameter~$\ell$ counts the number of vector generators
\be
G_i^{(n)} \qquad\textrm{with}\quad i=1,\ldots,d \quad\textrm{and}\quad n=0,\ldots,2\ell
\ee
which span the $\ell$-conformal Galilei algebra together with the $so(d)$ generators $M_{ij}$ for $1\le i<j\le d$
and the generators $P$, $D$ and~$K$ of the one-dimensional conformal algebra~\cite{confNH2,confNH3}.

Mechanical models invariant under $\ell $-conformal Galilei transformations typically contain higher time derivatives
for $\ell>\frac{1}{2}$ \cite{confNH3,confPU,GM1,Ivanov,MT,PU11,Mast}.
The method of nonlinear realizations~\cite{coset1,coset2} together with the inverse Higgs phenomenon~\cite{ih}
work quite well for the $\ell$-conformal Galilei algebra, giving rise to interesting invariant Lagrangians~\cite{GM1,Ivanov}.
However, recent results on the conformal invariance of Pais--Uhlenbeck oscillators~\cite{doa} with specific
frequencies~\cite{confPU, masterov}
appear to be add odds with the nonlinear-realization approach.
Indeed, in~\cite{confPU} the full $\ell$-conformal Galilei group was realized on a single bosonic field,
thus achieving conformal invariance for Pais--Uhlenbeck oscillators without a dilaton.
To describe such a situation within the nonlinear-realization approach, one has to include the dilatation~$D$
and the conformal boost~$K$ in the stability subgroup. Usually, such an extension of the stability subgroup is undesirable,
because the generators of the coset fail to form a representation of the stability subgroup,
and the construction of invariant actions becomes problematic.
In this paper we resolve this paradox by explicitly demonstrating (in Section~2) how actions can easily be constructed
for such unusual cosets and transformation properties of the fields and Cartan forms.
Moreover, we show that the minimal actions in $d{=}2$ describe Pais--Uhlenbeck oscillators with specific
frequencies\footnote{
These relations have been found in \cite{confPU, masterov}.}
\be
\omega, 3\omega, 5\omega, \ldots, 2\ell \omega \quad\textrm{for}\quad \ell\in\Z{+}\sfrac12 \und
2\omega, 4\omega, 6\omega, \ldots, 2\ell \omega \quad \textrm{for}\quad \ell\in\Z\ ,
\ee
where the basis frequency $\omega$ is a  parameter entering the nonlinear realization.

For $d{=}2$ it turns out to be advantageous to relabel the vector generators as~\cite{MT}
\be
G_\alpha \quad\textrm{and}\quad \bG_\alpha \qquad\textrm{with}\quad \alpha=-\ell,\ldots,\ell\ .
\ee
In this notation (see~\p{finalg1} below)
the $\ell{=}\frac12$-conformal Galilei algebra resembles the $N{=}2$ superconformal algebra in one dimension,
except that the fermionic generators are commuting (to zero).
This raises the question whether one can make these generators non-commuting,
thereby introducing a deformation of the Schr\"odinger algebra.
We analyze this possibility (in Section~3) and present the simplest invariant action describing
this newly deformed variant of the harmonic oscillator. We also find the general solution for its equation of motion.

As a byproduct of our deformation, the one-dimensional conformal group is realized in an unusual way on a single
complex bosonic field. We employ (in Section~4) our modified realization of the conformal group to generalize the
recent investigation~\cite{anton11, Anton12, Chernyavsky} of four-dimensional Ricci-flat metrics with SL(2,$\R$)~symmetry.
Reproducing their near-horizon Kerr metric for $\omega{=}0$, we find for $\omega{\neq}0$ a very specific modification
affecting only the radial and time variables. It is easy to include the effect of a cosmological constant,
which yields new constant-curvature metrics. In this case we present analytic expressions only for
a one-parameter family and for some isolated solutions.

\setcounter{equation}{0}
\section{Conformal Galilei group realization and Pais--Uhlenbeck oscillators}

Recently it has been shown \cite{confPU,masterov} that the Pais--Uhlenbeck oscillator enjoys an
$\ell$-conformal Newton--Hooke symmetry for half-integer or integer values of $\ell$ if the oscillation frequency
is an odd or even integer multiple of the basis frequency~$\omega$, up to $2\ell \omega$, respectively.
In this Section we are going to construct the minimal realization (on one complex bosonic field)
of $\ell$-conformal Galilei and Newton-Hooke symmetries for both integer and half-integer values of the parameter~$\ell$.
We will demonstrate that the simplest invariant actions describe the corresponding conformal Pais--Uhlenbeck oscillators.

To describe both, integer and half-integer, $\ell$-conformal symmetries we need to consider conformal Galilei symmetry
in $2{+}1$ dimensions where the corresponding algebra has the form
\bea\label{finalg1}
&& \im \left[ L_n, L_m \right] = (n-m) L_{n+m}, \quad n,m = -1,0,1, \quad \left( L_n\right)^\dagger = L_{n}, \nn \\
&&  \im \left[ L_n , G_\alpha \right] = ( n \ell - \alpha) G_{\alpha+n}, \quad \im \left[ L_n , \bG_\alpha \right] = ( n \ell - \alpha) \bG_{\alpha+n},
\quad \alpha= -\ell, -\ell+1, \ldots,\ell,
\; \left( G_\alpha\right)^\dagger = \bG_{\alpha}, \nn \\
&& \left[ U , G_\alpha  \right] = G_\alpha, \quad \left[ U , \bG_\alpha \right] = -\bG_\alpha, \quad \left( U\right)^\dagger =U.
\eea
This algebra admits a central-charge extension~\cite{confNH3} which looks slightly different for half-integer
and integer $\ell$,
\bea
\im \left[ G_\alpha, \bG_{\beta}\right] =
(-1)^{\ell+\alpha}\, (\ell+\alpha)!\, (\ell+\beta)!\,\delta_{\alpha+\beta,0}\, \hat{C},
&& \mbox{ for half integer $\ell, \alpha, \beta$}, \label{alg1} \\
\left[ G_\alpha, \bG_{\beta}\right] =
(-1)^{\ell+\alpha}\, (\ell+\alpha)!\, (\ell+\beta)!\,\delta_{\alpha+\beta,0}\, \hat{C},
&& \mbox{ for integer $\ell, \alpha, \beta$}, \label{alg2}
\eea
if we insist on the hermicity of the central charge ${\hat{C}}^\dagger = \hat{C}$.

We aim for a nonlinear realization of the $\ell$-conformal Galilei group $G$ in the coset $G/H$ with the choice of
\be\label{H0}
H\ = \ \mbox{span}( L_0, L_1, U,  \hat{C} ).
\ee
This choice of stability subgroup $H$ is quite unusual, because the dilatation $L_0$ and the conformal boost $L_1$
will then generate unbroken symmetries.  Previously~\cite{GM1}, nonlinear realizations of this group
took $L_0$ and $L_1$ to be spontaneously broken and thus always featured a dilaton among the physical fields.
In contrast, in our approach only $G_\alpha$ and $\bG_\alpha$ are spontaneously broken.

Our parametrization of the coset space reads
\be\label{g}
g= e^{\im t \left( L_{-1}+\omega^2 L_1 \right) } \prod_{\alpha=-\ell}^{\ell} e^{ \im \left( u_{\alpha} G_{\alpha} +\bu_{\alpha} \bG_{\alpha}\right)},
\ee
introducing a parameter~$\omega$.
Although~$\omega$ does not enter the $\ell$-conformal Galilei algebra, it affects its dynamical realization,
and in this context the term `$\ell$-conformal Newton-Hooke algebra' is often used~\cite{newton-hooke}.\footnote{
We thank A.~Galajinsky for pointing this out to us.}
The $\ell$-conformal Galilei group is realized by left multiplications on this coset,
\be
g_0 \, g = g' \, h,\, \qquad h \in H.
\ee
Thus, with respect to conformal transformations
$g_0=e^{\im ( a L_{-1}+ b L_0 +c L_1)}$ our fields $u_\alpha, \bu_\alpha$ and time $t$ transform as\footnote{
In what follows we will need to know the transformation properties of $u_{-\ell}$ and $\bu_{-\ell}$ only.}
\be\label{Conftr1}
\delta t = \frac{1+ \cos(2 \omega t)}{2}\, a + \frac{ \sin(2 \omega t)}{2 \omega}\, b +\frac{1-\cos(2 \omega t )}{2 \omega^2}\,c \equiv f(t), \quad
u'_{-\ell}(t') = \ell\, \dot{f} \, u_{-\ell}(t), \; \bu'_{-\ell}(t') = \ell\, \dot{f}\, \bu_{-\ell}(t).
\ee
Up to a redefinition of the parameters $a,b,c$, these transformations exactly coincide with those in~\cite{confPU}.

To find the transformation properties of the ``lowest-weight'' fields $u_{-\ell}$ and $\bu_{-\ell}$
under the shift symmetries generated by
\be\label{ShS}
g_\alpha = e^{\im ( b_\alpha G_\alpha +  {\bar b}_\alpha \bG_\alpha) },
\ee
one has to commute $g_\alpha$ past the factor $e^{\im t \left( L_{-1}+\omega^2 L_1 \right)}$ in the coset element~\p{g}.
This is achieved by employing the relation~\cite{BK1}
\be\label{aux1}
e^{\im t \left( L_{-1}+\omega^2 L_1 \right)} =
e^{\im \frac{\tan(\omega t)}{\omega} L_{-1}}\; e^{- 2 \im \log(\cos(\omega t)) L_0}\; e^{\im \omega \tan(\omega t) L_1},
\ee
which easily yields
\be\label{NHBtr}
\delta_\alpha u_{-\ell} = b_\alpha\; \frac{\tan^{\ell+\alpha}(\omega t)\, \cos^{2 \ell}(\omega t)}{\omega^{\ell+\alpha}}, \quad
\delta_\alpha \bu_{-\ell} = {\bar b}_\alpha\; \frac{\tan^{\ell+\alpha}(\omega t)\, \cos^{2 \ell}(\omega t)}{\omega^{\ell+\alpha}}.
\ee
Apparently, for any given value of $\ell$ the transformations \p{NHBtr} are just combinations of the shifts
\be
e^{- 2 \im\,\ell\omega t}, e^{- 2 \im\,(\ell-1)\omega t},\ldots, e^{ 2 \im\,(\ell-1)\omega t}, e^{ 2 \im\,\ell\omega t}
\ee
(including constants in the case of integer~$\ell$).

The next step is to calculate the Cartan forms for $g$ in \p{g},
\bea
&& g^{-1}\, d\, g =  \im \, \Omega_{-1} \left( L_{-1} +\omega^2 L_1\right) + \im \, \sum_{\alpha=-\ell}^\ell
\left( \omega_\alpha G_\alpha +{\bar\omega}_\alpha \bG_\alpha\right) +
\im \, \Omega_C \hat{C},  \nn \\
&& \Omega_{-1} = dt , \quad \omega_\alpha = du_\alpha+\cA_\alpha{}^\gamma\, u_\gamma\,dt ,  \quad {\bar\omega}_\alpha =  d\bu_\alpha+
\cA_\alpha{}^\gamma\, \bu_\gamma dt ,  \label{CF}
\eea
where $\cA_\alpha{}^\gamma$ is the $(2 \ell{+}1)\times (2 \ell{+}1)$ matrix
\be\label{A}
\cA = \left( \begin{array}{ccccccc}
0 & -1 & 0  & \ldots &0 &0 &0\\
2 \ell \omega^2 &0  & -2  & \ldots &0 &0 &0 \\
0 & (2 \ell{-}1) \omega^2 &0    & \ldots &0 &0 &0 \\
\vdots &\vdots &\vdots    & \ddots &\vdots &\vdots &\vdots\\
0 &0 &0    & \ldots & 0 &- 2\ell{+}1 & 0 \\
0 &0 &0    & \ldots & 2 \omega^2 &0 & -2 \ell \\
0 &0 &0   & \ldots &0 &\omega^2 &0
\end{array} \right).
\ee
The above calculations are quite similar to those ones performed in \cite{GM1}.

{}From the general theory of nonlinear realizations \cite{coset1, coset2} it follows that the  forms $\omega_\alpha$ and ${\bar\omega}_\alpha$  \p{CF}
are invariant with respect to the shift symmetries \p{ShS} and transform nontrivially under the conformal group \p{Conftr1}, because
\be\label{add1}
e^{\im ( a L_{-1} +b L_0 +c L_1)} \, g = g'\; e^{ \im ( b + c t ) L_0}\; e^{ \im c L_1}\; e^{ \im h {\hat{C}}}
\ee
and, therefore,
\bea
(g')^{-1} d g' & =& e^{ \im h {\hat{C}}} \; e^{ \im c L_1}\;e^{ \im ( b + c t ) L_0} \left( g^{-1} d g \right)
e^{ -\im ( b + c t ) L_0}\; e^{ -\im c L_1}\; e^{ -\im h {\hat{C}}} + \nn \\
&&\left( e^{ \im h {\hat{C}}}\; e^{ \im c L_1}\; e^{ \im ( b + c t ) L_0} \right) d \left(e^{ -\im ( b + c t ) L_0}\;
e^{ -\im c L_1}\; e^{ -\im h {\hat{C}}}\right).
\eea
The factor $e^{ \im ( b + c t ) L_0}$ just rescales the forms $\omega_\alpha$ \p{CF} and the factor $e^{ \im h {\hat{C}}}$ is harmless for these forms,
while the second factor $e^{ \im c L_1}$ will seriously reshuffle them. This is the price we have to pay for the non-orthonormal coset \p{g}.\footnote{
Orthonormality means that the coset generators form a representation of the stability subgroup.
In our case this is not so, because $\im [L_{-1}, L_1]=-2 L_0$.}
Nevertheless, the conditions \be\label{ih1}
\omega_\alpha = {\bar\omega}_\alpha= 0 \qquad\textrm{for}\quad \alpha=-\ell, \ldots, \ell{-}1
\ee
are invariant under all symmetries. Thus, the entire tower of fields
$(u_\alpha, \bu_\alpha | \alpha=-\ell{+}1,\ldots, \ell)$ may be covariantly
expressed through time derivatives of the lowest-weight fields $u_{-\ell}$ and $\bu_{-\ell}$ as
\bea
&& u_{-\ell+1} = \dot{u}{}_{-\ell}, \; u_{-\ell+2} = \sfrac{1}{2} \left( \ddot{u}{}_{-\ell} + 2 \ell \omega^2 u_{-\ell} \right),\nn\\
&&\bu_{-\ell+1} = \dot{\bu}{}_{-\ell}, \; \bu_{-\ell+2} = \sfrac{1}{2} \left( \ddot{\bu}{}_{-\ell} + 2 \ell \omega^2 \bu_{-\ell} \right),\quad
{\mbox{  etc.}} \label{ih2}
\eea
This is the inverse Higgs phenomenon \cite{ih}. In addition we impose the covariant constraints
\be\label{eom1}
\omega_\ell =  {\bar\omega}_\ell= 0 ,
\ee
which are just the equations of motion. It is not hard to check that they coincide with the
characteristic equation for the matrix $\cA$ \p{A} written for the time derivative $\frac{d}{dt}$, i.e.~with the equations
\bea
&&\left( \frac{d^2}{dt^2}+\omega^2\right)\left( \frac{d^2}{dt^2}+ 9 \omega^2\right)\ldots \left( \frac{d^2}{dt^2}+ (2 \ell)^2 \omega^2\right)u_{-\ell} =
\prod_{k=0}^{\ell-1/2} \left( \frac{d^2}{d t^2}+(2 k +1)^2 \omega^2 \right) u_{-\ell}=0,  \label{eom3}\\
&&\frac{d}{dt}\, \left( \frac{d^2}{dt^2}+4 \omega^2\right)\left( \frac{d^2}{dt^2}+ 16 \omega^2\right)\ldots \left( \frac{d^2}{dt^2}+ (2 \ell)^2 \omega^2\right)u_{-\ell} =
\prod_{k=1}^{\ell} \left( \frac{d^2}{d t^2}+ (2 k)^2 \omega^2 \right) {\dot u}_{-\ell} =0,   \label{eom4}
\eea
for half-integer and integer $\ell$, correspondingly.
Clearly, these equations follow from the conformally invariant Pais--Uhlenbeck oscillator actions
\be\label{Sol}
S_{\ell}^{PU} = \int dt \, \bu_{-\ell} \prod_{k=0}^{\ell-1/2} \left( \frac{d^2}{d t^2}+(2 k +1)^2 \omega^2 \right) u_{-\ell}
\quad\textrm{or}\quad
S_{\ell}^{PU} =  \im\,\int dt \,  \bu_{-\ell} \prod_{k=1}^{\ell} \left( \frac{d^2}{d t^2}+ (2k)^2 \omega^2 \right) {\dot u}_{-\ell},
\ee
respectively.

Thus, the nonlinear realization of the $\ell$-conformal Galilei group in the coset \p{g} gives rise to the conformally invariant Pais--Uhlenbeck oscillators.

We have checked in the lowest cases that the Lagrangian is just the Cartan form for the central charge $\hat{C}$.
Unfortunately, we did not succeed to bring the intermediate calculations into readable form and, hence,
a rigorous proof of this statement is lacking.

Let us complete this section with two comments.
\begin{itemize}
\item
The main difference of our nonlinear realization with those considered in \cite{GM1} is putting the generators $L_0$ and $L_{-1}$ into the stability subgroup~$H$~\p{H0}.
We may restore these generators via employing a coset parametrized by
\be\label{cosetIMP}
g= e^{\im t \left( L_{-1}+\omega^2 L_1 \right) } \prod_{\alpha=-\ell}^{\ell} e^{ \im \left( u_{\alpha} G_{\alpha} +\bu_{\alpha} \bG_{\alpha}\right)}
e^{\im u L_0} e^{\im z L_1}.
\ee
The additional factors will then seriously reshuffle the forms $\omega_\alpha, {\bar \omega}_\alpha$ of~\p{CF}.
Nevertheless, the full set of constraints $\omega_\alpha ={\bar\omega}_\alpha=0$ for $\alpha = -\ell, \ldots, \ell$
will produce the {\it same\/} set of equations of motion \p{eom3} and \p{eom4}. The dilaton $u$ will
decouple and obey the standard equation of motion upon imposing the additional constraints $\omega_{L_0}=\omega_{L_1}=0$ \cite{BK1}.
\item
However, if instead we use the coset
\be\label{cosetGM}
g= e^{\im t \left( L_{-1}+\omega^2 L_1 \right) } e^{\im u L_0} e^{\im z L_1} \prod_{\alpha=-\ell}^{\ell} e^{ \im \left( u_{\alpha} G_{\alpha} +\bu_{\alpha}
\bG_{\alpha}\right)}
\ee
as in~\cite{GM1}, then the equations of motion will get modified by interactions between the dilaton $u$ and the
fields $u_{-\ell}$ and $\bu_{-\ell}$. Passing from \p{cosetGM} to \p{cosetIMP} requires a redefinition of all the fields
$u_{\alpha}, \bu_{\alpha}$. In effect, we claim that the equations of motion of~\cite{GM1} may be decoupled
from the dilaton by a nonlinear redefinition of the fields.
\end{itemize}

\setcounter{equation}{0}
\section{A deformation of the Schr\"odinger algebra}
\subsection{Deformed oscillator}
The commutation relations of the $\ell$-conformal Galilei algebra written in the form \p{finalg1} are reminiscent of the relations of the wedge
subalgebra in
the Virasoro algebra extended by two commuting primary fields of the conformal weights $\ell{+}1$. {}From this analogy it is natural to ask:
Can one admit nontrivial relations between these primary fields,
i.e. make the shift generators $G_\alpha$ and $\bG_\alpha$ non-commuting?
A natural choice consists in the wedge subalgebra in some
nonlinear, W-type algebra discovered by A.B.~Zamolodchikov \cite{Zam1}.
Let us consider the simplest case of a such deformation.

The basic idea is to replace the $\ell=\frac{1}{2}$ conformal Galilei algebra by the factor algebra of the wedge subalgebra
in $W_3^{(2)}$ \cite{Polyakov, Bersh} over composite higher-spin generators, i.e.~a {\it linear\/} $su(1,2)$ algebra with the following commutation relations,
\bea\label{w32}
&&  \im \left[ L_n, L_m\right] = (n-m) L_{n+m}, \;  \im \left[L_n, G_r \right] =\left( \frac{n}{2}-r\right) G_{n+r}, \;\im
\left[L_n, \bG_r \right] =\left( \frac{n}{2}-r\right) \bG_{n+r}, \nn \\
&&  \left[U, G_r\right] =  G_r, \quad \left[U, \bG_r\right] =  - \bG_r,\nn\\[4pt]
&& \im \left[ G_r, \bG_s \right] = \gamma \left( \sfrac{3}{2} (r-s) U - \im  L_{r+s}\right), \quad n,m=-1,0,1,\; r,s = -1/2, 1/2.
\eea
Here, $\gamma$ is a deformation parameter:  if $\gamma=0$, we come back to the
$\ell=\frac{1}{2}$ conformal Galilei algebra. The exact value of $\gamma$ is inessential: if nonzero it can be put to unity by a rescaling of the generators $G_r$ and $\bG_r$.

We choose the stability subalgebra $H$ as
\be\label{H}
H\ =\ \mbox{span} ( L_0, L_1, U )
\ee
and realize this deformed symmetry by left multiplications of
\be\label{coset33}
g=e^{\im t \left( L_{-1}+\omega^2 L_1\right)}\, e^{\im \left(u G_{-1/2}+\bu \bG_{-1/2}\right) }\,  e^{\im \left(v G_{1/2}+\bv \bG_{1/2}\right)}.
\ee
The transformation properties of the time $t$ and the lowest-weight fields $u ,\bu$ get deformed for $\gamma\neq0$:
\bea\label{tr34}
g_0=e^{\im \,a L_{-1}}: & & \left\{
\begin{array}{l}
\delta t = a \left( \sin^2(\omega t)+ \frac{4 \cos(2 \omega t)}{4 - \gamma^2 \omega^2 (u\, \bu)^2} \right), \;
\delta u  =-\frac{a}{2} \omega u \left(  \sin( 2 \omega t) + \frac{4 \im \, \gamma \omega \cos(2 \omega t)}{4- \gamma^2 \omega^2 (u\, \bu)^2} u\, \bu\right),
\end{array}\right.  \nn \\
g_0=e^{\im \, b L_0} : && \left\{
\begin{array}{l}
\delta t = \frac{b \sin(2 \omega t)}{2 \omega} \left(  \frac{4 + \gamma^2 \omega^2 (u \, \bu)^2}{4 - \gamma^2 \omega^2 (u \, \bu)^2} \right),\;
\delta u  =\frac{b}{2}  u \left( \cos( 2 \omega t) - \frac{4 \im \gamma \omega \sin(2 \omega t)}{4- \gamma^2 \omega^2 (u\, \bu)^2} u\, \bu\right),
\end{array}\right. \nn \\
g_0=e^{\im \,c L_1}: &&\left\{
\begin{array}{l}
\delta t = \frac{c}{\omega^2} \left( \cos^2(\omega t)- \frac{4 \cos(2 \omega t)}{4- \gamma^2 \omega^2 (u\, \bu)^2} \right),\;
\delta u  =\frac{c}{2 \omega}  u \left(  \sin( 2 \omega t) + \frac{4 \im \gamma \omega \cos(2 \omega t)}{4- \gamma^2 \omega^2 (u\, \bu)^2} u\, \bu\right),
\end{array}\right.  \nn \\
g_0=e^{\im \left( a G_{-1/2}+ {\bar a} {\bar G}_{-1/2}\right)}: &&  \left\{
\begin{array}{l}
\delta t = \frac{  2\im \gamma \cos(\omega t\,) ({\bar a} u - a \bu) + \gamma^2 \omega \sin(\omega t) ({\bar a} u + a \bu)\, u\, \bu }
{4- \gamma^2 \omega^2 (u\, \bu)^2} \\
\delta u  =a \cos(\omega t) -\frac{\im \gamma \omega}{2} \sin(\omega t)\, u (2 {\bar a} u + a \bu) -\frac{\im}{2} \gamma \omega^2 u^2\, \bu \delta t,
\end{array}\right.  \nn \\
g_0=e^{\im \left( b G_{1/2}+{\bar b} \bG_{1/2}\right) }: && \left\{
\begin{array}{l}
\delta t = \frac{ 2 \im \gamma \sin(\omega t)\, ({\bar b} u - b \bu) - \gamma^2 \omega \cos(\omega t) ({\bar b} u + b \bu) \, u\,\bu }{\omega\,(4- \gamma^2 \omega^2 (u\, \bu)^2)} \\
\delta u  = \frac{\sin(\omega t)}{\omega} b  +\frac{\im \gamma}{2} \cos(\omega t)\, u (2 {\bar b} u + b \bu) -\frac{\im}{2} \gamma \omega^2 u^2 \, \bu \delta t,
\end{array}\right.  \nn \\
g_0=e^{\im \alpha U}: && \delta u = \im \alpha u.
\eea
In what follows, we will need only the Cartan forms $\omega_{\pm 1/2},{\bar\omega}_{\pm 1/2}$ and $\omega_U$ which read
\bea\label{CF34}
\omega_{-1/2} & = &d u + \sfrac{\im}{2}\gamma\,\omega^2 u^2 \, \bu\, dt - v d\tau,\quad
{\bar\omega}_{-1/2}  = \left(\omega_{-1/2}\right)^*, \nn \\
\omega_{1/2} & = &d v + \sfrac{\im}{2}\gamma\,v^2\, \bv\, d\tau -\sfrac{\im}{2}\gamma\,v\,\left[ 2 v \left(d \bu  -
\sfrac{\im}{2}\gamma\,\omega^2 u \, \bu^2\, dt\right) +
\bv \left(d u + \sfrac{\im}{2}\gamma\,\omega^2 u^2\, \bu\, dt\right)\right]  \nn \\
&& +\ \sfrac{3\im}{2}\gamma \,\omega^2 v\, u\, \bu\, dt + \omega^2\, u\, dt, \quad
{\bar\omega}_{1/2}  = \left( \omega_{1/2}\right)^*, \nn \\
\omega_U & = & \sfrac{3}{2} \gamma \left[ v\, \bv\, d\tau - v \left( d \bu  -\sfrac{\im}{2}\gamma\,\omega^2 u \, \bu^2\, dt\right) -
\bv\, \left(d u + \sfrac{\im}{2}\gamma\,\omega^2 u^2 \, \bu\, dt\right) +\omega^2 u\, \bu\,dt \right], 
\eea
where
\be
d\tau =  \left( 1+ \sfrac14 \gamma^2 \, \omega^2 u^2\, \bu{}^2\right) dt+ \sfrac{\im}{2}\gamma\left( u\, d \bu - \bu\, du\right).
\ee
The inverse Higgs constraints are the same as in the undeformed case,
\be\label{ih34}
\omega_{-1/2}={\bar\omega}_{-1/2}=0\qquad \Rightarrow \qquad v = \frac{\dot{u}+\im \frac{\gamma \, \omega^2}{2} u^2\, \bu}{1+\im \frac{\gamma}{2}
\left( u \dot{\bu} - \bu \dot{u}\right) +\frac{\gamma^2 \, \omega^2}{4} u^2\, \bu^2},\;
\bv = \frac{\dot{\bu}-\im \frac{\gamma \, \omega^2}{2} u\, \bu{}^2}{1+\im \frac{\gamma}{2} \left( u \dot{\bu} - \bu \dot{u}\right) +
\frac{\gamma^2 \, \omega^2}{4} u^2\, \bu^2}.
\ee
With these constraints taken into account, the form $\omega_U$ simplifies to
\be\label{UU}
\omega_U = -\sfrac{3}{2} \gamma \left( v\, \bv\, d\tau - \omega^2 u\, \bu\, dt\right).
\ee
Observing that under all transformations \p{tr34} the form $\omega_U$ only shifts by an exact differential,
we can write down a simple invariant action,
\be\label{lag3}
S=  -\sfrac{2}{3 \gamma} \int \omega_U =
 \int dt \frac{\dot{u}\, \dot{\bu} - \omega^2 u\, \bu }{1+\im \frac{\gamma}{2}( u\, \dot{\bu} - \bu\, \dot{u})+\frac{1}{4} \gamma^2 \omega^2 u^2 \bu^2}.
\ee
The equations of motion following from this action coincide with those obtained from the constraints
\be\label{eomm}
\omega_{1/2} = {\bar\omega}_{1/2}=0\qquad \Rightarrow\qquad \dot{v}- \im \gamma v^2\,
\left(\dot{\bu} -\sfrac{\im}{2} \gamma \, \omega^2 u \, \bu^2\right) +\omega^2 u \left(  \sfrac{3 \im}{2} \gamma v \bu +1\right)=0,
\ee
where $v, \bv$  are defined in \p{ih34}.

We conclude that the deformation of the symmetry algebra, i.e.~the passing from the $\ell=1/2$ Galilei algebra to the $su(1,2)$ algebra produces a non-polynomial velocity dependence in the action~\p{lag3}.
The ``free''($\omega=0$) system shares this feature. The undeformed ($\gamma=0$) case describes a harmonic oscillator
(or, with $\omega=0$, a free particle). The intriguing question is whether our deformation
preserves the integrability of the harmonic oscillator? In the next section we will prove this by explicit construction of the general solution for the system  \p{lag3}.

\subsection{General solution}
The integrability of the system \p{lag3} is evident due to conservation of the angular momentum
\be\label{i3}
 \im \left( \bu v - u \bv \right) +\gamma u \bu v \bv = \mbox{const} =: C ,
\ee
which commutes with the Hamiltonian\footnote{
The standard expression for the Hamiltonian may be obtained by using
$p = \left( 1+ \frac{\im \gamma}{2} \bu v\right) \bv$ and $\bp =\left( 1 - \frac{\im \gamma}{2}  u \bv\right) v$.}
\be\label{Hf}
H= \left( 1+ \sfrac14 \omega^2 \gamma^2 u^2 \bu{}^2 \right) v \bv -\sfrac{\im}{2} \omega^2 \gamma\,u \bu \left( u \bv -\bu v\right)+ \omega^2 u \bu .
\ee
Thus, one may directly solve the equations of motion.

However, the deformed oscillator \p{lag3} possesses some interesting properties which allow us to find the general solution in a purely
algebraic way. Let us summarize these properties.
\begin{itemize}
\item The currents $I$ and $\bar I$ obey oscillator equations,
\be\label{eqi12}
I=\bv \left( 1 + \im \gamma \bu v\right)\;,\quad  {\bar I} =v \left( 1 - \im \gamma u \bv\right)
\qquad\Rightarrow\qquad
\ddot{I}+\omega^2 I=0 \und \ddot{\bar I}+\omega^2 {\bar I}=0.
\ee
\item The current $I_2$ oscillates with twice the frequency,
\be\label{eqi4}
I_2 = u \bv + \bu v \quad \Rightarrow \quad \ddot{I_2}+ 4 \omega^2 I_2 =0.
\ee
\item Employing the evident solutions of \p{eqi12} and \p{eqi4},
\be\label{eq}
I= A \sin(\omega t)+ B \cos(\omega t), \; {\bar I} = {\bar A} \sin(\omega t)+ {\bar B} \cos(\omega t),\quad
I_2 = \mu \frac{ \sin(2 \omega t)}{2 \omega} + \nu \cos(2 \omega t),
\ee
and the conservation \p{i3} one finds algebraically that
\be\label{ssol}
u = \frac{ 2 \im (1+\gamma C) +\gamma I_2 - \im \sqrt{4 (1+\gamma C) - \gamma^2 I_2^2}}{2 \gamma I} \und
v= \frac{\left( \im \gamma I_2 +\sqrt{4 (1+\gamma C) - \gamma^2 I_2^2}\right) {\bar I}}{2 (1 +\gamma C)}.
\ee
\item The constraints \p{ih34} yield three restrictions on the coefficients $(A, {\bar A}, B, {\bar B}, \mu, \nu)$ in \p{eq},
which may be solved for $A$, $\bar A$ and~$\mu$,
\be\label{fsol}
\mu= \frac {2 B {\bar B }}{1 +\gamma C}-\omega^2 \frac{(\nu^2+\omega C^2)(1+\gamma C)}{2 B {\bar B}},\quad
A=-\omega \frac{(\nu-\im C)(1+\gamma C)}{2 {\bar B}},\;
{\bar A}=-\omega \frac{(\nu+\im C)(1+\gamma C)}{2 B},
\ee
reducing the independent constants to $(B,{\bar B},\nu,C)$, which is the anticipated number.
\end{itemize}
The energy of this solution is given by the Hamiltonian $H$ \p{Hf} as
\be
E= \frac { B {\bar B }}{1 +\gamma C}+\omega^2 \frac{(\nu^2+ C^2)(1+\gamma C)}{4 B {\bar B}} =\frac{1}{2} \left( \mu + \frac{4 A {\bar A}}{1+\gamma C}\right).
\ee
We note that, for the special value $C= -\frac{1}{\gamma}$, the system has only the trivial solution $u=\bu=0$.

\subsection{The $\omega{=}0$ case}
It is worth commenting on the $\omega=0$ case. In this limit everything greatly simplifies. The action reads
\be\label{lagm}
S_0=  \int dt\ \frac{\dot{u}\, \dot{\bu} }{1+\im \frac{\gamma}{2}( u\, \dot{\bu} - \bu\, \dot{u})},
\ee
while its symmetry transformations acquire the form
\bea\label{tr0}
g_0=e^{\im \,a L_{-1}}: & & \left\{
\begin{array}{l}
\delta t = a ,
\end{array}\right.  \nn \\
g_0=e^{\im \, b L_0} : && \left\{
\begin{array}{l}
\delta t = b\, t ,\;
\delta u  =\frac{b}{2}  u ,
\end{array}\right. \nn \\
g_0=e^{\im \,c L_1}: &&\left\{
\begin{array}{l}
\delta t = c t^2 -\frac{1}{4} c \gamma^2 \left( u \, \bu \right)^2,\;
\delta u  =c t u  +\frac{\im}{2} c \gamma u^2 \bu,
\end{array}\right.  \nn \\
g_0=e^{\im \left( a G_{-1/2}+ {\bar a} {\bar G}_{-1/2}\right)}: &&  \left\{
\begin{array}{l}
\delta t = \frac{\im}{2} \gamma \left( {\bar a} u - a \bu \right),
\delta u  =a ,
\end{array}\right.  \nn \\
g_0=e^{\im \left( b G_{1/2}+{\bar b} \bG_{1/2}\right) }: && \left\{
\begin{array}{l}
\delta t = \frac{\im}{2} \gamma t \left({\bar b} u - b \bu\right) - \frac{1}{4} \gamma^2 u \bu \left( {\bar b} u + b \bu\right)  \\
\delta u  = b\, t +\frac{\im}{2} \gamma u \left( 2\, {\bar b}\, u + b\, \bu \right),
\end{array}\right.  \nn \\
g_0=e^{\im \alpha U}: && \delta u = \im \alpha u.
\eea
The expressions for the higher Goldstone fields $v$ and $\bv$ \p{ih34} become very simple,
\be\label{ih0}
v = \frac{\dot{u}}{1+\im \frac{\gamma}{2} \left( u \dot{\bu} - \bu \dot{u}\right)},\;
\bv = \frac{\dot{\bu}}{1+\im \frac{\gamma}{2} \left( u \dot{\bu} - \bu \dot{u}\right)},
\ee
as do the equations of motion,
\be\label{eom0}
\dot{v}- \im \gamma v^2\,  \dot{\bu} =0.
\ee
Since for $\omega=0$ the Hamiltonian $H$ \p{Hf} reduces to
\be\label{H00}
H_0=  v \bv ,
\ee
that the equations \p{eom0} merely state that the currents
\be\label{i120}
I=\bv \left( 1 + \im \gamma \bu v\right) \und {\bar I} =v \left( 1 - \im \gamma u \bv\right)
\ee
are conserved,
\be\label{eqi120}
\dot{I}=0 \und \dot{\bar I}=0.
\ee
Thus, these currents -- corresponding to the shift generators $G_{-1/2}$ and $\bG_{-1/2}$ --
commute with the Hamiltonian \p{H00}.

Concerning the general solution of the equations of motion \p{eom0}, we still have a free equation for the current $I_2$ \p{eqi4},
\be\label{eqi40}
I_2 = u \bv + \bu v \quad \Rightarrow \quad \ddot{I_2}=0,
\ee
and therefore obtain
\be\label{eq0}
I= B , \; {\bar I} = {\bar B},\quad I_2 = \mu \, t  + \nu .
\ee
The angular momentum \p{i3} is still conserved,
and so one may algebraically find the solution of the equations of motion in the form
\be\label{ssol0}
u = \frac{ 2 \im (1+\gamma C) +\gamma I_2 - \im \sqrt{4 (1+\gamma C) - \gamma^2 I_2^2}}{2 \gamma I}.
\ee
Again, checking the relations \p{ih0} determines the coefficient $\mu$ in \p{eq0} to
\be\label{fsol0 }
\mu= \frac {2 B {\bar B }}{1 +\gamma C}.
\ee
The energy of this solution is derived from the Hamiltonian $H_0$ \p{H00},
\be
E= \frac { B {\bar B }}{1 +\gamma C} =\sfrac{1}{2} \, \mu .
\ee

\setcounter{equation}{0}
\section{Massive extension of near-horizon Einstein metrics}
\subsection{Deforming the near-horizon Kerr metric}
In the $\omega=0$ limit, considered in the previous subsection, the transformations of our fields $u, \bu$ and the time $t$ exhibit interesting properties.
After passing to a new basis $\{ t, \rho, \phi \}$ defined by
\be\label{newcoor}
\rho = \frac{1}{u\,\bu}, \quad \phi = -\frac{\im}{2} \log( \frac{u}{\bu}),
\ee
the $so(1,2)$ transformations \p{tr0} acquire the form
\be\label{newtr}
\delta t = a + b t + c\left( t^2 - \frac{\gamma^2}{4 \rho^2}\right), \;  \delta \rho = - b \rho - 2 c t \rho, \;  \delta \phi = \frac{c \gamma}{ 2 \rho},
\ee
where, as before, the parameters $a,b,c$ correspond to translations $(L_{-1})$, dilatations $(L_0)$ and conformal boosts $(L_1)$, respectively.
This form resembles the realization of the one-dimensional conformal symmetry found by Bardeen and Horowitz \cite{BH}
in considering the near-horizon limit of the four-dimensional Kerr black hole.\footnote{
The precise relation is obtained via replacing
$\gamma \rightarrow 2 \im \gamma$ and $\phi\rightarrow - 2 \im \phi$.}

In the same basis \p{newcoor}, the $\omega\neq 0$  $so(1,2)$ transformations \p{tr34} read
\bea\label{newtr1}
&& \delta t = a \left(\frac{ 4 \rho^2 \cos( 2 \omega t)}{ 4 \rho^2 - \omega^2 \gamma^2}+\sin^2(\omega t)\right) +
b \frac{(4 \rho^2 +\omega^2 \gamma^2)\sin(2 \omega t)}{2 \omega (4 \rho^2 -\omega^2 \gamma^2 )}
- \frac{c}{\omega^2} \left( \frac{ 4 \rho^2 \cos( 2 \omega t)}{ 4 \rho^2 - \omega^2
\gamma^2}-\cos^2(\omega t)\right), \nn \\
&& \delta \rho = a \omega \sin( 2 \omega t) \rho  - b \cos(2 \omega t) \rho -c \frac{ \sin (2 \omega t) \rho}{\omega}, \nn \\
&& \delta \phi = - a \frac{2 \omega^2 \gamma \cos( 2 \omega t) \rho}{ 4 \rho^2 - \omega^2 \gamma^2} -b \frac{ 2 \omega \gamma \sin(2 \omega t) \rho}{4 \rho^2 - \omega^2 \rho^2}+
c  \frac{2  \gamma \cos( 2 \omega t) \rho}{ 4 \rho^2 - \omega^2 \gamma^2} .
\eea
Does there exist some $\omega\neq 0$ deformation of the near-horizon Kerr solution in which the conformal $so(1,2)$ symmetry
is realized as in \p{newtr1}? To answer this question we apply the procedure performed in~\cite{Anton12}.
The conformal invariants entering the near-horizon metric read
\be\label{inv}
\omega_1 = \left( \frac{2\rho}{\gamma}-\omega^2\frac{\gamma}{2\rho}\right)^2 dt^2 +\frac{ d\rho^2}{\rho^2}, \quad
\omega_2 = \left( \frac{2\rho}{\gamma}+\omega^2\frac{\gamma}{2\rho}\right) dt + 2 d\phi, \quad d \theta,
\ee
where $\theta$ is the latitudinal angular variable, which is inert under the conformal transformations \p{newtr1}.
With these invariants
one may express the most general four-dimensional conformally invariant metric as
\be\label{metric0}
\begin{aligned}
ds^2 &= & F(\theta) \left[ \left( \frac{2\rho}{\gamma}-\omega^2\frac{\gamma}{2\rho}\right)^2 dt^2
+\frac{d\rho^2}{\rho^2}+d\theta^2 \right] -
G(\theta) \left[ \left( \frac{2\rho}{\gamma}+\omega^2\frac{\gamma}{2\rho}\right) dt + 2 d\phi \right]^2 .
\end{aligned}
\ee
Note that shifting $\rho\to \omega\rho$ and $t\to \omega^{-1}t$ corresponds to putting $\omega=1$, and
redefining $\rho\to\gamma\rho$ amounts to setting $\gamma=1$.
The vacuum Einstein equations
\be\label{einstein}
R_{\mu\nu}=0
\ee
impose conditions merely on the coefficient functions $F(\theta)$ and $G(\theta)$:
\be\label{eqmetrics}
2 (F + F'')(2 F -F'') + 3 (F'+F^{\prime\prime\prime})\,F' =0 \und
G= -\frac{ (F')^2}{F} + \frac{4}{3}( F + F'') .
\ee
Somewhat surprisingly, the mass parameter $\omega$ does not enter these equations, and thus they are identical to
the $\omega=0$ case studied in~\cite{Anton12}. Referring to
this paper for a detailed analysis, we reproduce here the general solution,
\be\label{newsol}
F(\theta) = C_1 \left( 1 +\cos^2 (\theta)\right)+ C_2 \cos (\theta)
\qquad\Rightarrow\qquad
G(\theta) = \frac{\left( 4 C_1^2-C_2^2\right)\,\sin^2(\theta)}{C_1 \left( 1 +\cos^2 (\theta)\right)+ C_2 \cos (\theta)},
\ee
with arbitrary integration constants $C_1$ and $C_2$.
The third integration constant hides in a trivial constant shift of $\theta$.
It is remarkable that the solution space is linear (i.e.~it admits superpositions) although the equation is not.
We conclude that the modified realization \p{tr34} of the conformal $so(1,2)$ symmetry introduces a ``frequency'' modification into the four-dimensional Ricci-flat metrics constructed (for $\omega=0$) in~\cite{Anton12}.
The physical interpretation of the metric \p{metric0} with the solution \p{newsol} requires passing back
to Minkowski signature via $\gamma \rightarrow -\frac{\im}{2} \gamma$ and $\phi\rightarrow \frac{\im}{2} \phi$
and remains an open challenge.

\subsection{Deforming the near-horizon Kerr--dS/AdS metric}
It is easy to extend the construction to general Einstein metrics, i.e.~constant-curvature metrics,
by adding a cosmological constant~$\Lambda$ to~\p{einstein}. Demanding
\be\label{Lambda}
R_{\mu \nu} + \Lambda g_{\mu \nu} =0
\ee
for the metric~\p{metric0} changes the conditions~\p{eqmetrics} to
\bea\label{eqmetrics1}
&& 2 (F + F'')(2 F -F'') + 3 (F'+F^{\prime\prime\prime})\,F'-2 \Lambda F\left( 4 F^2 +3 (F')^2 + F F''\right)+ 4 \Lambda^2 F^4 =0, \\ \label{eqmetrics2}
&& G= -\frac{ (F')^2}{F} + \frac43( F + F''-\Lambda F^2) ,
\eea
which is no longer homogeneous under rescaling of~$F$.
However, the $\Lambda$ dependence may be absorbed in a rescaling $F\to F/\Lambda$.
It follows that solutions blow up in the $\Lambda\to 0$ limit, unless their overall scale is variable.
The equation for~$F$ can be rewritten as
\be\label{ad1}
\frac{1}{F'} \left( F+F''-\Lambda \, F^2\right)^2
\frac{d}{d\theta} \left[ 4 F - \frac{3\,(F')^2}{F+F''-\Lambda \, F^2}\right] =0,
\ee
and so it reduces to solving\footnote{
The solution $F = C_0$ to equation \p{addeq} is admitted only for $C_0=1/\Lambda$ as is seen from \p{eqmetrics1}.}
\be\label{addeq}
4 F - \frac{3\,(F')^2}{F+F''-\Lambda \, F^2} = 4\,C_0
\qquad\Rightarrow\qquad
4 \left( F - C_0 \right) \left(F+F''-\Lambda F^2\right) - 3 (F')^2=0
\ee
with some integration constant~$C_0$.
In the limit of $C_0\to\infty$ we are led to
\be\label{addeq2}
F+F''-\Lambda F^2 =0,
\ee
so this possibility of solving \p{ad1} is already included in~\p{addeq}.
The general solution to \p{addeq} contains
one more integration constant besides~$C_0$ and the trivial $\theta$ shift.
In the limit $\Lambda\to0$, we can compare with the general solution~\p{newsol} in the pervious subsection
and confirm this count by the explicit relation
\be
4\,C_1(C_1-C_0) = C_2^2 \qquad\textrm{for}\quad \Lambda=0.
\ee
The full explicit solution to \p{eqmetrics1} is given by elliptic functions and will not be displayed here.
We would like to mention four special cases however.
First, for $C_0=3/4\Lambda$ we have the particular solution
\be\label{newsol3}
F(\theta)= \frac{3}{2 \Lambda \left(1 + \cos (\theta)\right)}
\qquad\Rightarrow\qquad
G(\theta) = -\frac{3\,\sin^2(\theta)}{\Lambda \left(1 + \cos (\theta)\right)^3}.
\ee
The other three cases occur at $C_0\to\infty$, i.e.~are solutions to~\p{addeq2}. Second,
\be\label{newsol2}
F(\theta)= \frac{1}{\Lambda}+\frac{144\,e^{\theta}}{(1 - 24 \Lambda e^{\theta})^2}
\qquad\Rightarrow\qquad
G(\theta) =- \frac{20736\Lambda e^{2\theta} (1+24 \Lambda e^{\theta})^2}
{(1-24 \Lambda e^{\theta})^4 (1+96 \Lambda e^{\theta}+576 \Lambda ^2 e^{2\theta})} .
\ee
Third, there is the trivial constant solution
\be\label{newsol0}
F(\theta)= \frac{1}{ \Lambda}
\qquad\Rightarrow\qquad
G(\theta) = 0
\ee
to be discarded since it produces a singular metric.
Fourth and most interesting, the $C_1=0$ family of \p{newsol} smoothly extends to a family of
$\Lambda\neq0$ solutions,
\be\label{newsol1}
F(\theta) =\frac{ C_2 \left(  \cos (\theta) +\frac{1}{3} \Lambda C_2\right)}
{\left(1  +\frac{1}{3}\Lambda C_2  \cos(\theta)\right)^2}
\qquad\Rightarrow\qquad
G(\theta) = -\frac{27 C_2 \sin^2(\theta)}{(3 \cos(\theta)+\Lambda\, C_2)\left(3+\Lambda\, C_2 \cos(\theta)\right)^2},
\ee
since the naturally singular $1/\Lambda$ behavior could be absorbed into~$C_2$ in this case.

\setcounter{equation}{0}
\section{An extended Niederer transformation}
It should be clear from our construction that it is possible to redefine the time $t$ and the fields $u, \bu, v, \bv$ which parameterize our coset
\p{coset33} to bring it into an $\omega$-independent form,
\be\label{NT1}
g=e^{\im t \left( L_{-1}+\omega^2 L_1\right)}\, e^{\im \left(u G_{-1/2}+\bu \bG_{-1/2}\right) }\,  e^{\im \left(v G_{1/2}+\bv \bG_{1/2}\right)} \qquad\rightarrow\qquad
{\tilde g}= e^{\im \tau L_{-1}}\, e^{\im \left(\tilde{u} G_{-1/2}+\tilde{\bu} \bG_{-1/2}\right) }\,  e^{\im \left(\tilde{v} G_{1/2}+\tilde{\bv} \bG_{1/2}\right)}.
\ee
The exact relation between two bases reads
\bea\label{NT2}
\tau &= & \frac{4 -\omega^2 \gamma^2 u^2 \bu{}^2}{4+ \omega^2\, \gamma^2\, \tan^2 (\omega t)\,  u^2\, \bu{}^2} \;\left( \frac{\tan ( \omega t)}{\omega}\right), \nn \\
\tilde u & = & \frac{ 2 \im u}{ 2 \,\im + \omega \,\gamma\, \tan (\omega t)\,  u \,\bu}\;\left( \frac{1}{\cos (\omega t)}\right), \quad
\tilde v = \frac{\frac{d}{d \tau} {\tilde u}}{1+\im \frac{\gamma}{2}
\left( \tilde{u}\, \frac{d}{d \tau}{\tilde\bu} - \tilde{\bu}\, \frac{d}{d \tau}{\tilde u}\right)}.
\eea
This is an extended version of the Niederer transformation \cite{NTr} which maps a harmonic oscillator to a free particle.
One may easily check that the action  \p{lag3} and the metric \p{metric0} acquire the form
\bea\label{Ntr3}
S & =&\int dt \frac{\dot{u}\, \dot{\bu} - \omega^2 u\, \bu }{1+\im \frac{\gamma}{2}( u\, \dot{\bu} - \bu\, \dot{u})+\frac{1}{4} \gamma^2 \omega^2 u^2 \bu^2} =
\int d\tau \frac{ \frac{d}{d \tau}{\tilde u}\, \frac{d}{d \tau}{\tilde \bu}}{1+ \im \frac{\gamma}{2} \left( \tilde{u} \frac{d}{d \tau}\tilde{\bu} -
\tilde{\bu}\frac{d}{d \tau}{\tilde u}\right)} , \nn \\[4pt]
ds^2 &= & F(\theta) \left[ \left( \frac{2\rho}{\gamma}-\omega^2\frac{\gamma}{2\rho}\right)^2 dt^2
+\frac{d\rho^2}{\rho^2}+d\theta^2 \right] -
G(\theta) \left[ \left( \frac{2\rho}{\gamma}+\omega^2\frac{\gamma}{2\rho}\right) dt + 2 d\phi \right]^2 \nn \\
&= & F(\theta) \left[ \frac{4\tilde{\rho}^2}{\gamma^2} d\tau^2 +\frac{d\tilde{\rho}^2}{\tilde{\rho}^2}+d\theta^2 \right] -
G(\theta) \left[ \frac{2\tilde{\rho}}{\gamma} d\tau + 2 d\tilde{\phi} \right]^2,
\eea
where
\be
\tilde{\rho} = \frac{1}{\tilde{u}\,\tilde{\bu}}, \quad \tilde{\phi} = -\frac{\im}{2} \log( \frac{\tilde{u}}{\tilde{\bu}}).
\ee
Due to the known features of the Niederer transformation, the equivalence between the two theories is only a local one.

\setcounter{equation}{0}
\section{Conclusions}
We have constructed a minimal field realization (on a single complex boson) of the
$\ell$-conformal Galilei group in $2{+}1$ dimensions.
The simplest actions, given by the integral of the Cartan form for the central-charge generator, describe
conformally invariant Pais--Uhlenbeck oscillators.

The main difference between our approach and previous ones~\cite{GM1} is in the structure
of the stability subgroup $H$, in which we put the generators of dilatation and conformal boost.
Despite the non-orthonormality of the resulting coset space we could construct covariant equations of motion
by imposing proper restrictions on the Cartan forms.

We have found it useful to employ a special basis for the $\ell$-conformal Galilei algebra,
in which the shift generators resemble primary spin-$(\ell{+}1)$ fields of a Virasoro algebra.
This basis has also been used, for example, in~\cite{MT}.
This analogy may be prolonged further by deforming the conformal Galilei algebra to include a wedge subalgebra
in some nonlinear $W$-type algebra.
We did this for the simplest $\ell{=}1/2$ Galilei algebra, i.e.~for the Schr\"odinger algebra.
We constructed the simplest action for this case and proved that the corresponding system
and its $\omega \neq 0$ extension are both integrable and solvable.

Concerning further developments, we make the following remarks.
\begin{itemize}
\item Our deformation is not limited to the case we presented.
For example, one may contemplate an $su(n)$ deformation of the Schr\"odinger algebra,
based on the wedge subalgebra in a quasi-superconformal algebra (see e.g.~\cite{BG} and references therein).
One may also consider a deformation of the $\ell=1$ conformal Galilei algebra
as well as the deformations for other values of $\ell$.
\item To shed light on the interpretation of the proposed deformation, it is important to look at
the quantum deformed oscillator and its spectrum.
\item One may investigate the supersymmetric extension of the deformed oscillator,
which should be based on the wedge subalgebra in the ${\cal N}{=}2$ super $W_3^{(2)}$ algebra \cite{N2w32}.
\item Along the line we proposed, deformations of other conformal Galilei algebras will also yield
modified transformation laws for the time parameter and the fields, thus providing novel realizations
of the $d=1$ conformal algebra. The latter may be employed for constructing new four-dimensional
Einstein metrics along the line of~\cite{anton11,Chernyavsky}.
\end{itemize}

\section*{Acknowledgements}
We are grateful to Anton Galajinsky and Armen Nersessian for valuable correspondence.
The work of S.K. was partially supported by RSCF grant 14-11-00598 and by
RFBR grant 15-52-05022 Arm-a.
The work of O.L. and of A.S. was partially supported by DFG grant Le-838/12-2.
The work of A.S. was partially supported also by
RFBR grants 15-52-05022 Arm-a and 16-52-12012-a and the Heisenberg-Landau program.
This article is based upon work from COST Action MP1405 QSPACE,
supported by COST (European Cooperation in Science and Technology).

\end{document}